# Distributed scientific communication in the European information society: Some cases of "Mode 2" fields of research

Gaston Heimeriks*
Loet Leydesdorff**
Peter Van den Besselaar*

*Department of Social Science Informatics (S.W.I), Universiteit van Amsterdam, Roetersstraat 15
1018 WB  Amsterdam, The Netherlands

** Science & Technology Dynamics, Universiteit van Amsterdam, Nieuwe Achtergracht 166, 1018 WV Amsterdam, The Netherlands

# Abstract

Can self-organization of scientific communication be specified by using literature-based indicators? In this study, we explore this question by applying entropy measures to typical "Mode-2" fields of knowledge production. We hypothesized these scientific systems to be developing from a self-organization of the interaction between cognitive and institutional levels: European subsidized research programs aim at creating an institutional network, while a cognitive reorganization is continuously ongoing at the scientific field level. The results indicate that the European system develops towards a stable level of distribution of cited references and title-words among the European member states. We suggested that this distribution could be a property of the emerging European system. In order to measure to degree of specialization with respect to the respective distributions of countries, cited references and title words, the mutual information among the three frequency distributions was calculated. The so-called transmission values informed us that the European system shows increasing levels of differentiation.

# 1 Introduction

In the project 'Self-organization of the European Information Society'[1] new methodologies of combining theoretical specification with quantitative modeling are developed to stimulate the understanding of the complex social processes that characterize the development of the information society. Can the formation of a European identity be retrieved in scientometric data? Can self-organization of scientific communication be specified by using literature-based indicators? In this study, we explore this question by limiting ourselves to typical "Mode-2" fields of scientific research. Fields of techno-science like biotechnology, artificial intelligence (AI) and information sciences develop in a network mode: disciplinary insights from different backgrounds are recombined and university-industry relations are continuously reshaped. The ongoing process of integration at the European level generates an additional network of transnational collaborations. Using the co-occurrences of title words and cited references of scientific publications in core journals of biotechnology, AI and information sciences, entropy analysis can be used to distinguish between the intellectual organization of the publications and the institutional structure in terms of addresses of documents. This analysis focuses on the grouping of co-occurrences of cited references and title-words in various levels of aggregation. The question is whether countries and the EU are appropriate levels of aggregation to study specialization patterns in mode 2 sciences. In a first set of analysis the overlap of research topics in various geographic groups is determined: the global system is decomposed into Japan, USA and EU, the European system is decomposed into the member states. In a more elaborate study the specialization patterns are determined with respect to the same groupings. A straightforward way to measure to degree of specialization with respect to countries and research topics is to calculate the mutual information among the three frequency distributions of countries, cited references and title words. The so-called transmission values are a measure of dependence between distributions and originated from information theory as a measure of transmission.

---

[1] The authors acknowledge funding by the European Commission, TSER-project SOE1-DT97-1060, "The Self-Organization of the European Information Society."

Entropy statistics and information theory are tools for analyzing complex, distributed systems. The concept of entropy is used here to study the distribution of cited references and title words on the aggregated national level. Though the concept of entropy originated from thermodynamic systems, it has acquired a general probabilistic meaning that allows for a large number of applications (Theil, 1969, 1972; Langton, 1990). Entropy statistics is based solely on the properties of probability distributions, and, as such, is especially suitable for studying evolutionary phenomena at the level of any population of heterogeneous entities (Saviotti, 1996). This applies also to evolving communication systems.

Social systems consist of distributed communication, which can be studied by plotting communications in a matrix of interactions (Luhmann, 1982; Leydesdorff, 1993). In this way, action with reference to an individual can be considered communication with reference to the system. The network is spanned in terms of relations, but it develops a specific architecture, in which each action also has a position. The Science Citation Index provides us with a wealth of data to plot communication patterns in science. In case of the knowledge producing communication system, scientists are the carrying nodes. Central to this model is the idea that science can be characterized by the distributed communication of 'papers', each proposing a new quantum of knowledge (Gilbert, 1997). The actors carrying the network are 'authors' who write papers. The production of papers involves a number of scientometric attributes that all play an important role in scientific communication (and all show the typical skewed distribution of self-organized processes). By producing a scientific communication the author performs a number of actions; the selection of a journal, the choice of title-words (representing a topic), and the choice of cited references indicating a context in which the research takes place. Furthermore, authors can select other authors to co-write papers. All the actions performed by scientists are reflexive, motivated by increasing the chances of getting the paper published or the patent accepted.

The development of the resulting communication system is expected to behave in a self-organizing mode. The theory of self-organization applies to this operationally closed system. In the case of the European information society, two major processes of reconstruction can be distinguished. First, there is the networking of the European Union in terms of institutional addresses. Among other things, the series of Framework Programs and other special programs have stimulated both transnational collaboration and collaboration across sectors.

In addition to this level of networking, the new techno-sciences like biotechnology, new materials, and information technologies contain another network dynamics, namely, at the cognitive and therefore potentially global level. Gibbons et al. (1994) have coined the word "Mode-2 research" for areas like biotechnology. The applicational contexts are considered constitutive of the development of these fields. From the global perspective, the European policies construct an intermediate network layer among some of the competing agencies.

# 2 Methodology and Data

### 2.1. Delineation of the domain

Scientific specialties are expected to behave as self-organizing communication systems. The main mechanism of self-organization is recursion of selection, that is, non-randomness of the events. Apparently, some kind of 'plastic identity' (Maturana, 1980) is maintained by the system through time, since the name and key concepts of the fields under study are remaining identical. The most frequently cited references are exemplary for this phenomenon of plastic identity[2]. Consequently, the systems have internal operations to maintain the preconditions of self-organization: operational closure, complex structure, high dynamics and feedback mechanisms.

---

[2] To illustrate this point: The top ten of most cited references in the scientific specialty 'biotechnology' remain surprisingly constant in the ten-year period under study; 7 out of 10 cited references are identical in 1996 and 1986.

Of course, despite the maintained identity, the definition of the fields under study does change. If one follows the actors historically (Latour 1987), one obtains a reflexive understanding about how these definitions have changed. From an evolutionary perspective, however, one expects that some of these historical elements have been carried over into the current understanding, while other elements may have faded away.

A method for empirical analysis of evolving scientific communication systems has been developed that reflects these phenomena of systems boundaries and identity construction. Using aggregated journal-journal citations as listed in the Journal Citation Reports (JCR) of the Science Citation Index (SCI), the citational environment of the entrance journal is defined as all journals that cite or are cited by this journal above a threshold percentage. The method is described in Cozzens and Leydesdorff (1993) and Van den Besselaar & Leydesdorff (1996). Factor analysis reveals clearly identifiable clusters of journals representing scientific disciplines. By repeating this procedure using data from previous years, information is obtained about the recent development of the field and its environment. The data of all journals in the clusters under study were obtained from the SCI.

In studying the evolving knowledge producing system, we limit ourselves to the techno-sciences biotechnology, information science and artificial intelligence (AI). These disciplines were chosen not only because of their dynamic developments in an applicational context, but also because of efforts of European subsidized research programs to establish trans-national institutional networks in these fields.

The evolutionary perspective can be obtained by focusing on the operation of the system in the present[3]. On the basis of previous research, we conjectured that Biotechnology and Bioengineering could be considered as a leading journal in this field of biotechnology (Leydesdorff & Gauthier 1996). Using this seed journal as an entrance to the Journal Citation Reports of the Science Citation Index for 1996, a relevant journal environment was delineated by taking all journals into consideration that cited this journal or were being cited by it to the extent of one percent of its total citations. These aggregated citation relations were organized in a matrix, which was then factor analyzed. Since most cells are empty and the dense clusters of citation represent scientific specialties, factor analysis resulted in clearly identifiable clusters.

The eigenvector representing "biotechnology" consists of the thirteen journals, among which Biotechnology and Bioengineering not only has the highest impact factor of the set; it also has the highest factor loading on the factor indicating biotechnology. It can therefore be considered as a

---

[3] For reasons of accessibility of data, the 'starting-point' for analysis was 1996.

**4**

"central tendency journal" in the citing dimension, as defined by Cozzens & Leydesdorff (1993). With hindsight, this legitimates the initial selection of Biotechnology and Bioengineering as the starting point for the analysis. Data of the same set of journals was obtained for the years 1992 and 1986 in addition to the starting year 1996[4].

Information science was delineated using the *Journal of the American Society for Information Science* (JASIS) as entrance journals at a 1 % threshold. Again, we created from the JCR-1996 the Journal-Journal citation matrix. Factor-analyzing this matrix resulted in six factors, of which the first factor is the one around JASIS. The factor explains 35% of total variance. The second factor is a library science factor with four journals, explaining about 19% of total variance. Apart from these two factors, we identified four smaller ones in 1996. Again, 1992 and 1986 completed the evolutionary period of analysis.

The delineation analysis for artificial intelligence (AI) is described in Van den Besselaar and Leydesdorff 1996. The journal Artificial Intelligence served as an entrance journal, again the factor containing the entrance journal explained the largest part of the total variance. In the case of AI 1988, 1992 and 1997 were used for analysis.

**2.2. Methodological design**

Entropy measures are used to identify sub-dynamics and characteristics of emerging science systems in the information society. The method of applying entropy measures to distributed communication systems is useful because entropy measures are flexible with respect to the level of aggregation, thus comparisons between different levels are possible (Theil, 1972, Leydesdorff, 1995). We compared a European system, composed of 15 EU-member states with the global system composed of the EU, Japan, and U.S.A. Of course, there are many more countries in the world than these three industrial blocks, but we know from previous research that including more countries is not always helpful from an analytical perspective. For example, including Canada makes it more difficult to distinguish between the U.S.A. and Europe because of this country's position between the U.S. and the U.K. research systems (Leydesdorff & Gauthier 1996). In this study, we analyze various Mode 2 specialties at different levels of aggregation in order to better understand the possible emergence of a specific European level.

In the previous section we described how the papers were obtained that together define a scientific specialty. All data from the SCI could be organized in a relational database allowing for reorganization of the records as the basis for the institutional analysis. In addition to an institutional address, the papers provide a variety of attributes like a list of cited references and a title. Different frequency lists of title-words and cited references could be retrieved from all papers with an institutional address in the EU, Japan or U.S.A respectively. This data was used to construct matrices of co-occurrences of the most cited references (± 250) and the most occurring title-words (± 250) per country and per region as represented in figure 1. The global system is decomposed into the EU, Japan and U.S.A. The European system is decomposed in its 15 member states.

---

[4] In a later stage, 1997 was added to the analysis because 1992 produced deviating values.

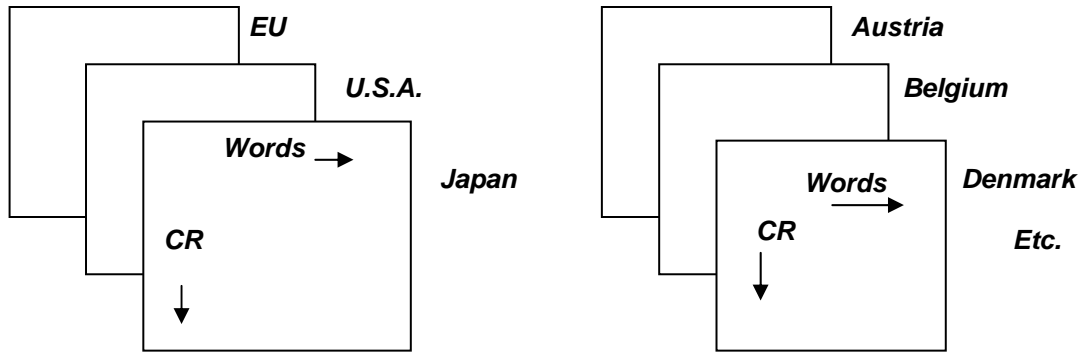

*Figure 1. The organization of the scientometric data for analysis in matrices that are composed of co-occurrences of most cited references (CR) and most frequently used title-words (Words). On the left the global system decomposed in terms of the EU, Japan and U.S.A. The European system decomposed into 15 member states is represented on the right*

In the model above, we represented scientific topics as a set of co-occurrences of words and references. We distinguish two levels of aggregation. The first one, is the global system of Japanese, American and European co-occurrences and secondly, the European system of all the EU countries. The entropy value of each two-dimensional matrix is given by;

$$H(A) = - \sum_{a=A} (p_a \cdot \log_2 p_a)$$

Entropy is a measure of variance or uncertainty: the higher the entropy, the more difficult it is to predict the scientific topics as indicated by the use of cited references and title words which is blindly picked from the population. The entropy is zero when all distributions are equal since then there is no uncertainty, and is positive otherwise. The larger the entropy value, the larger the variety within a distribution of topics.

We will decompose the co-occurrences of cited references and title-words at the system level in terms of contributions of each subgroup. The $H_0$ is then the between-group entropy, which is the difference between the higher-level entropy and the sum of the weighted lower level entropies. In this way, the value of $H_0$ is a measure for the quality of the 'grouping'.

$$H(A)_{total} = H(A)_0 + \sum_{a=A} \left( p_a \cdot H_a \right)$$

In our analysis, the $H_{tot}$ is the entropy measure on the aggregate level, e.g. the EU or the global level. The last term is the weighted sum of all countries' values; the average within-group entropy. The difference between the two, the $H_0$ is a measure of dividedness (Theil, 1972). The total entropy is always larger than the between-group entropy, since there cannot be more dividedness after grouping than there was before grouping.

## 2.2 Transmission values

In the next step we applied entropy measures to the distribution of co-occurrences in order to determine the transmission values, or mutual information between the frequency distributions. The distribution of co-occurrences in the three-dimensional space of words-references-countries is not homogeneous but highly skewed. (figure 2).

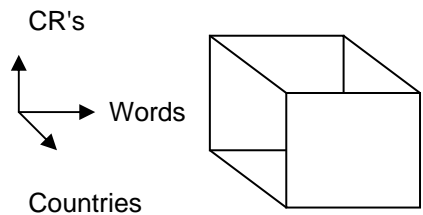

*Figure 2. The 'cube' of co-occurrences of words, cited references and countries to which entropy measures are applied.*

The aim of this analysis is to quantify the developments of specialization with respect to the different levels of aggregation; the EU vis a vis Japan and USA, and the EU members state vis a vis eachother. A straightforward way to measure to degree of specialization is to calculate the mutual information among the three frequency distributions (Frenken, 2000). The transmission values are a measure of dependence between distributions and originated from information theory as a measure of transmission. (Theil 1967, 1972; Leydesdorff, 1995).[5] The three-dimensional specialization measure however, is only a summary indicator, the specialization may occur in only two of the three dimensions. We represented scientific topics as a set of co-occurrences of words and references. Again, we distinguish two levels of aggregation. The first one, is the global system of Japanese, American and European co-occurrences and secondly, the European system of all the EU countries. To detect in which pairs of dimensions the specialization patterns are most pronounced we will study the transmissions between the pairs of cited references and words, cited references and countries and words and countries. The cited references are represented along the X-axis, the title words on the Y-axis. The co-occurrences of cited references and title-words are distributed in the dimension of geographical groups, the Z-axis. When the respective entropies in three dimensions are calculated it follows that (Theil, 1972; Leydesdorff, 1995; Frenken, 2000);

$T_{xy} = H_x + H_y - H_{xy}$
$T_{xz} = H_x + H_z - H_{xz}$
$T_{yz} = H_y + H_z - H_{yz}$
$T_{xyz} = H_x + H_y + H_z - H_{xyz}$

In sum, entropy can be used as a measure of observational variety. Entropies are a function of relative magnitudes, probabilities being the most common form. The sample size does not influence the entropy values (Krippendorff). The question is how we can analyze the distribution of co-occurrences of title-words, cited references and corporate sources. The words and cited references can be used as indicators for of intellectual organization, while the corporate sources are known ex ante. Since the knowledge producing system is a complex communication system, we expect self-organization to occur, e.g. a non-random, non-homogeneous distribution to occur. In order to retrieve the formation of a European scientific system, we shall apply entropy analysis to the three industrial blocks of the E.U., the U.S.A., and Japan, and then to the 15 European member states.

## 3. Results

### 3.1 Quality of the grouping

---

[5] See Frenken (2000) for an excellent and detailed description of the mathematical operationalization of this methodology.

The first step in our analysis of distributed scientific communication was to determine to which extend patterns of cited references and title words are similar among different countries, or in other words, to which extend are the grouping variables 'country' differentiated? The data concerning the co-occurrences of the most frequently used title words and cited references aggregated on nation level (the EU countries) and on block-level (Japan, USA and EU) were organized in matrices as shown above. The entropy values of the individual matrices and the overall set were calculated and the between group entropy values ($H_0$) could then be determined. These values are a measure for the 'dividedness' of patterns of co-occurrences between countries (Theil, 1972). The $H_0$ values shown in table 1 are normalized with the entropy value of the overall co-occurrences matrix. In this way, the European and global $H_0$'s are comparable.

Table 1. Results of entropy analysis of between group values.

|  | Htot Global | $\Sigma H$ | H0 | %H0 | Htot EU | $\Sigma H$ | H0 | %H0 |
|---|---|---|---|---|---|---|---|---|
| **AI** | | | | | | | | |
| **1988** | 10.82 | 10.04 | 0.78 | 7.21 | 8.18 | 6.21 | 1.97 | 24.08 |
| **1992** | 11.2 | 10.16 | 1.04 | 9.29 | 9.81 | 6.87 | 2.94 | 29.97 |
| **1997** | 11.16 | 10.05 | 1.11 | 9.95 | 10.33 | 7.4 | 2.94 | 28.46 |
| **INFO** | | | | | | | | |
| **1986** | 10.54 | 9.72 | 0.82 | 7.76 | 8.7 | 6.95 | 1.75 | 20.07 |
| **1992** | 11.38 | 10.44 | 0.94 | 8.26 | 10.16 | 7.85 | 2.31 | 22.77 |
| **1996** | 11.62 | 10.62 | 1.00 | 8.64 | 10.55 | 7.71 | 2.84 | 26.89 |
| **BIO** | | | | | | | | |
| **1986** | 11.56 | 10.22 | 1.34 | 11.59 | 10.74 | 7.84 | 2.90 | 27.05 |
| **1992** | 11.87 | 10.64 | 1.24 | 10.36 | 11.40 | 8.32 | 3.08 | 27.01 |
| **1996** | 12.67 | 11.44 | 1.23 | 9.71 | 12.23 | 8.81 | 3.42 | 27.96 |

The between-group entropy ($H_0$) is defined as the difference between the entropy prevailing at the lower level ($\Sigma H$) and the entropy at the higher level ($H_{tot}$). To interpret these values, note that the total entropy is equal to the between-group entropy plus the average within group entropy. It follows that a relative high between-group value can be interpreted as a relative different distribution of co-occurrences in the different groups in the matrix of events, suggesting a differentiated communication system, while an identical distribution would result in a $H_0$ of zero (Theil, 1972).

In AI the entropy values show a rise in $H_0$ values, especially between 1988 and 1992, indicating a development of specialization with respect to the geographical origin of the co-occurrences of cited references and title words in this period. The European system follows the same trend as the global system. In information science a similar trend can be observed; a steady rise of dividedness between 1986 and 1996 on the global and European level.

Most striking is the difference in the level of differentiation on the global level between biotechnology on one hand, and AI and info-science on the other hand. In biotechnology we see a decrease in differentiation in the period 1986 to 1996, while the total number of published documents and co-occurrences rises in that period. In information science and AI the level of differentiation appears to approach a stable value, yet starting from a lower level than that in biotechnology. However, within the European knowledge producing system the three disciplines develop from different levels to a stable and identical level of differentiation among the European member states. Might this distribution indicate a property of the European system?

## 3.2 Degree of specialization

The analyses to determine the transmission values were motivated by the idea that the co-occurrences of cited-references and title-words do not occur homogeneously. Does the level of aggregation 'EU' gain in significance or not, and the European member state as level of aggregation? The entropy value of the distribution of co-occurrences over countries can be calculated, starting from the distribution of co-occurrences in the three-dimensional space $H_{xyz}$. The H values will inform us that about the overall increase in variety that occurred in the period under study. The overall transmission values (T xyz) will inform us about the level of specialization within Europe and on the global level. As mentioned before, the three-dimensional specialization measure however, is only a summary indicator. Specialization patterns may occur in only two dimensions.

|   |   | AI |   | INFO |   | BIO |   |
|---|---|---|---|---|---|---|---|
|   |   | 88 |   | 86 |   | 86 |   |
| sigma H |   | 10.04 | 6.21 | 9.72 | 6.95 | 10.22 | 7.84 |
| H xyz | H cube | **10.82** | **8.18** | **10.54** | **8.70** | **11.56** | **10.74** |
| H xy | H matrix | **10.72** | **8.17** | **10.37** | **8.53** | **11.32** | **10.40** |
| H xz | H cr-countries | **7.98** | **6.41** | **7.46** | **6.37** | **8.15** | **8.21** |
| H yz | H w-countries | **6.95** | **5.19** | **6.41** | **5.73** | **8.00** | **8.35** |
| H x | H CR | **7.61** | **6.30** | **7.01** | **5.86** | **7.27** | **6.86** |
| H y | H Words | **6.59** | **4.90** | **5.87** | **5.04** | **7.06** | **6.83** |
| H z | H countries | **0.77** | **1.97** | **0.82** | **1.75** | **1.35** | **2.90** |
| T xy | TRANSMISSION | 3.47 | 3.03 | 2.51 | 2.37 | 3.01 | 3.29 |
| T xz |   | 0.40 | 1.86 | 0.37 | 1.24 | 0.46 | 1.55 |
| T yz |   | 0.41 | 1.68 | 0.28 | 1.05 | 0.40 | 1.38 |
| **Txyz** |   | 4.15 | 4.99 | 3.16 | 3.95 | 4.12 | 5.85 |
|   |   | 92 |   | 92 |   | 92 |   |
| sigma H |   | 10.16 | 6.87 | 10.44 | 7.85 | 10.67 | 8.32 |
| H xyz | H cube | **11.20** | **9.81** | **11.38** | **10.16** | **11.87** | **11.40** |
| H xy | H matrix | **11.12** | **9.61** | **11.28** | **9.98** | **11.70** | **11.20** |
| H xz | H cr-countries | **8.40** | **7.57** | **7.89** | **7.35** | **8.45** | **8.45** |
| H yz | H w-countries | **6.99** | **7.57** | **7.25** | **7.24** | **8.04** | **8.59** |
| H x | H CR | **7.72** | **6.97** | **7.46** | **6.67** | **7.84** | **7.57** |
| H y | H Words | **6.45** | **5.70** | **6.62** | **6.43** | **7.13** | **6.98** |

| H z | H countries | **1.04** | **2.94** | **0.94** | **2.32** | **1.21** | **3.08** |
|---|---|---|---|---|---|---|---|
| T xy | TRANSMISSION | 3.05 | 3.06 | 2.80 | 3.12 | 3.27 | 3.36 |
| T xz | | 0.37 | 2.33 | 0.51 | 1.63 | 0.60 | 2.20 |
| T yz | | 0.50 | 1.07 | 0.31 | 1.51 | 0.30 | 1.48 |
| Txyz | | 4.01 | 5.80 | 3.64 | 5.25 | 4.30 | 6.23 |
| | | 97 | | 96 | | 96 | |
| sigma H | | 10.05 | 7.40 | 10.62 | 7.71 | 11.44 | 8.81 |
| H xyz | H cube | **11.16** | **10.33** | **11.62** | **10.55** | **12.67** | **12.23** |
| H xy | H matrix | **10.90** | **10.03** | **11.48** | **10.41** | **12.51** | **11.95** |
| H xz | H cr-countries | **7.96** | **7.91** | **8.39** | **8.05** | **8.79** | **9.19** |
| H yz | H w-countries | **7.69** | **7.75** | **7.11** | **7.30** | **8.39** | **9.46** |
| H x | H CR | **7.22** | **6.86** | **7.90** | **7.39** | **8.02** | **7.79** |
| H y | H Words | **6.93** | **6.51** | **6.43** | **6.22** | **7.36** | **7.29** |
| H z | H countries | **1.12** | **2.94** | **1.00** | **2.84** | **1.23** | **3.42** |
| T xy | TRANSMISSION | 3.25 | 3.34 | 2.85 | 3.20 | 2.86 | 3.14 |
| T xz | | 0.37 | 1.88 | 0.52 | 2.18 | 0.46 | 2.02 |
| T yz | | 0.36 | 1.69 | 0.32 | 1.76 | 0.20 | 1.25 |
| **Txyz** | | 4.11 | 5.98 | 3.72 | 5.89 | 3.94 | 6.27 |

In general, we observe an increase in overall specialization on the European level as indicated by the Txyz value, indicating differentiation between European countries. In AI and biotechnology, there is no clear indication of specialization on the global level. In information sciences however, an increase in global Txyz values can be observed on the global level as well as on the European level.

The two dimensional indicators are expected to inform us on the underlying characteristics of these developments. In AI we observe unusual characteristics on the global level; the 1988, 1992 and 1997 the transmission value Tyz is larger than the Txz. This means that in the global AI system, the words are more specific with respect to the geographical origin of a scientific paper than the cited references. In Europe, as well as in biotechnology and information sciences (on EU and global level), the cited references are more specific with respect to the geographical origin.

The developments in information sciences can be further elaborated by studying the two-dimensional transmission values. The results show that the two dimensional transmission values increase in all cases. The rise is strongest in the dimensions with cited references however, in T xz and T xy. It seems that a strong increase in the variety of cited references has occurred. Consequently, this variety of references allows for more unique country-reference combinations and more unique combinations of cited references and title words.

What we see in biotechnology is interesting; the overall differentiation on the global level decreases. Homogenization seems to take place on the global level. At the European level however, an increase in specialization can be observed. The mutual information between references and words remains the same at European level, and decreases lightly on the global level. However, the mutual information content between the dimensions words and countries decreased stronger, while the mutual information content between cited references and countries remains the same on the global level. Especially this development of Tyz on the global level is interesting, it seems that words become decreasingly differentiated with respect to their geographical origin.

The European biotechnology system differentiates. The transmission of the dimensions words and references shows a stable value. The differentiation of the European biotechnology systems seems to be the result of increasingly differentiated combinations of countries and references. While countries share the same vocabulary in European biotechnology studies, the way researchers draw on earlier works, is differentiating.

## 4 Scientometric consequences

In general, countries are more differentiated with respect to the cited references than with respect to the title-words. The only exception is the global AI system. In biotechnology, the increased level of specialization within the European Union can be contributed to an increase in specialization with respect to the choice of cited references. The use of words became less specific with respect to geographical origin. Scientometric indicators like cited references and title-words can be distinguished by the way they refer to different levels of the textual organization. From an evolutionary perspective on scientific communication, the text –that is words and co-words- provide the variation (Callon, 1986). By referencing, a subset of these texts is selected. References and citations can therefore be considered indicators of these selection processes (Leydesdorff and Wouters, 1999).

A possible explanation for the exceptional a-specificity of cited references with respect to the geographical origin of the document in the global level AI system, could be the small size of the discipline compared to biotechnology and information sciences.

The way researchers draw on earlier works, and their sharing of a set of exemplars, is considered to be reflected in the referencing practices of the specialty members. On the other hand, the shared interest in a set of research problems and concepts is expected to be reflected in the word-patterns. The congruence in both mapping practices is presupposed in most scientometric studies. In a previous study this assumption was criticized by Braam et al. (1991). We provided a quantification for the relationship between the indicators cited references and title-words, by means of the transmission values. The two dimensional transmission value $T_{xy}$ is a measure for the overlap between the frequency distributions of references and title-words. Although the $T_{xy}$ values do not show a constant level of mutual information between the two dimensions, Txy seems to approach a constant value of around 3 bits indicating that topics are generally defined by a partially overlapping group of cited references and title-words. The extend to which the two dimensions overlap might be related to the stage of paradigmatic development of a specialty. The mutual information content of cited references and title-words increases, in parallel with the previous observation of the development of the distribution of research topics in the European system (table 1). Thus, while the distribution of research topics seems to develop towards a stable level among the European member-states, the mutual information content of the indicators cited references and title-words that together define a topic, seems to develop towards a stable level as well. The results indicate that the European system develops towards a stable level of distribution among the European member states. This distribution could be a property of the emerging European system, because it is independent from the number of papers in a discipline. The development towards a stable distribution of topics seems to be related with the paradigmatic state of the respective scientific disciplines.

## 5. Concluding remarks

The study was motivated by the hypothesis that the self-organization of the European Information society is the result of two co-evolving developments. Transnational collaborations are a prerequisite for European funding thus an institutional European network is likely to occur. On the other hand the cognitive development is beyond control of policy makers and scientists. The entropy analyses informed us that an increasing differentiation has occurred in the European science systems under study. The level of differentiation between the EU, Japan and USA however, remained the same (a small increase can be observed in information sciences). From the above stated hypothesis it follows that it is unlikely to find strong differentiation on the global level. The cognitive 'definition' of a field takes place on this level of aggregation. The European science system shows increasing levels of differentiation. However, all three disciplines show considerable growth in the period under study. The increase in European specialization can be the result of an increase in the number of contributions to the fields under study; in this way the variety of co-occurrences increases. The differentiation can also be the result of an increase in the number of countries that substantially contribute to the fields under study in patterns that show (partial) overlap with existing specialization patterns.

In order to interpret these results, it might be necessary to study the nature of the above mentioned developments in more detail. In the next part of this study, a more qualitative approach will be adopted. In this model, one expects that countries in a transnational collaboration recombine their individual national specialization patterns. When an increase in the number of

publications –and consequently an increase in co-occurrences- takes place, it is likely that transnational collaboration will result in partially overlapping transnational research efforts. In the next section, the co-occurrence and cited references will be grouped in research topics; the entropy results seem to support this methodology. The mutual information between word patterns and patterns of references approaches a stable value, suggesting an underlying structure of research topics. The contribution of the various countries to a small number of research topics can be studied by visualizing the respective contributions.